\documentstyle[aps,prb,epsfig]{revtex}

\begin{document}

\title{Origin of combination frequencies in quantum magnetization oscillations of two-dimensional multiband metals}
\author{Thierry Champel}
\address{
Commissariat \`{a} l'Energie Atomique,  DRFMC/SPSMS\\
17 rue des Martyrs, 38054 Grenoble Cedex 9, France}
\date{\today}

\maketitle

\begin{abstract}
The influence of chemical potential oscillations on the magnetization 
oscillations in two-dimensional  multiband metals is investigated.
In a first part, the analytical derivation of Alexandrov and 
Bratkovsky [Phys. Rev. B \textbf{63}, 033105 (2001)] 
of magnetic quantum oscillations in two-dimensional multiband metals with a fixed number of electrons is shown to be mathematically incorrect ;
the chemical potential oscillations appearing in the arguments of the Fourier components were not taken into account.
In a second part, we derive an approximative Fourier 
series of the magnetization oscillations in the regime of small chemical potential oscillations.
 The main result is that combination 
frequencies with significant amplitudes are found if the individual band 
frequencies differ significantly.

\end{abstract}


The observation of quantum magnetization oscillations (the de Haas-van Alphen effect) is
one of many manifestations of the quantization of the electronic 
energy spectrum in a magnetic field $H$. This quantum effect is observable 
both in two-dimensional (2D) and three-dimensional (3D) metals. 
However, on account of the importance of chemical potential oscillations, the de
 Haas-van Alphen (dHvA) effect in 2D metals is quite different from its 3D 
counterpart. It becomes dependent on experimental conditions : the 
magnetization oscillations show different characteristics corresponding to 
either imposing a constant number of electrons (for which the chemical potential $\mu$ 
oscillates with the magnetic field) or to imposing a fixed chemical 
potential~\cite{1,2,3}.
For example, in 2D multiband metals, unusual combination frequencies of the 
ordinary frequencies of the individual bands appear in the Fourier spectrum 
of magnetization oscillations in the presence of chemical potential oscillations. 
The observation~\cite{4} of these additional contributions to the dHvA effect 
with significant amplitudes enables a quantitative comparison to theory to be made.
For this purpose, an analytical description of the dHvA effect in 2D multiband 
metals including chemical potential oscillations on a quantitative basis 
is needed.

Recently, Alexandrov and Bratkovsky~\cite{5} have proposed an analytical 
derivation for the amplitudes of the combination frequencies in magnetic 
quantum oscillations. They argued that, for a fixed number of electrons $N$, 
the combination frequencies originate from the oscillation of a squared term, only present in the expression 
of the free energy $F$ in the canonical ensemble. After calculating the 
grand canonical thermodynamic potential $\Omega$, they obtained the explicit 
form of the free energy [Eq. (13) of Ref. 5] using the relation 
$F=\Omega+ \mu N$. 
From their expression (13), they got straightforwardly the Fourier harmonics 
of the combination frequencies.

According to Ref. 5,  the field oscillating part $\tilde{F}$ of the free 
energy $F$ is related to the field oscillating part $\tilde{\Omega}$ of the 
grand canonical thermodynamic potential $\Omega$ through

\begin{equation}
\tilde{F}=\tilde{\Omega}-\frac{1}{2 \rho} \left( \frac{\partial \tilde{\Omega}}{\partial \mu} \right)_{H}^{2},
\end{equation}
where $\rho$ is the total density of states.
Following Alexandrov and Bratkovsky~\cite{5}, it is the second nonlinear 
oscillatory term in Eq. (1), specific to the canonical ensemble, which 
produces the combination frequencies.
However, it is important to stress here that $\tilde{\Omega}$ is an explicit 
function of the variable $\mu$  with the ratio $\mu/\omega_{c \alpha}$ appearing in the arguments of the Fourier components
(where $\omega_{c \alpha}$ is the cyclotron frequency of the individual band 
$\alpha$), see explicit form Eq. (6), Ref. 5.
 For a fixed total number of electrons $N$, the correct treatment is thus to 
eliminate the variable $\mu$ in the expression for $\tilde{\Omega}$ 
via the condition 
\begin{equation}
N= - \left( \frac{\partial \Omega}{\partial \mu} \right)_{H}.
\end{equation}
Under a magnetic field, the chemical potential consists of a constant part 
$\mu_{0}$ (the value at zero magnetic field) plus an oscillating part 
$\tilde{\mu}$ given by the implicit equation

\begin{equation}
\tilde{\mu}=\frac{1}{\rho} \left( \frac{\partial \tilde{\Omega}}{\partial \mu} \right)_{H}.
\end{equation}

In 3D metals, the chemical potential oscillations are negligibly 
small~\cite{6} since $\tilde{\mu} \sim \omega_{c} \sqrt{\omega_{c}/\mu_{0}}$ 
and $\mu_{0} \gg \omega_{c}$ (in this part, the band index $\alpha$ is 
omitted for convenience). The oscillating part $\tilde{\mu}$ appears in the arguments of the Fourier components of $\tilde{\Omega}$ in the combination
$\tilde{\mu}/\omega_{c} \sim \sqrt{\omega_{c}/\mu_{0}}$.
Therefore, the direct substitution of the variable $\mu$ by the constant 
part $\mu_{0}$ in the explicit expression of $\tilde{\Omega}$ is 
valid~\cite{6} : $\tilde{\Omega}(\mu) \approx \tilde{\Omega}(\mu_{0})$. It 
means that the exact resolution of implicit Eq. (3) is not needed. In this 
case, the explicit expression for $\tilde{\Omega}$ can effectively be 
considered as a Fourier series~\cite{7}.
Moreover, the second term of Eq. (1), 
$\left( \partial \tilde{\Omega}/\partial \mu \right)_{H}^{2}/2 \rho \sim \rho \omega_{c}^{2} \left(\omega_{c}/\mu_{0}\right),$ 
is negligible compared with $\tilde{\Omega} \sim \rho \omega_{c}^{2} \sqrt{\omega_{c}/\mu_{0}}$. 
In 3D multiband metals, Eq. (1) is therefore applicable but combinations frequencies arising from the squared term in expression (1) have very small amplitudes.

In 2D multiband metals and at very low temperature $T \ll \omega_{c}$, the 
situation is different.
As expected in these conditions, the amplitude of chemical potential 
oscillations is much greater than in 3D metals. 
According to Eq. (13) of Ref. 5, 
$\tilde{\Omega} \sim \rho \omega_{c}^{2} \sim \left( \partial \tilde{\Omega}/\partial \mu \right)_{H}^{2}/2 \rho$ 
at low temperature and for a weak impurity scattering : the two terms in 
expression (1) are then of the same order.
However, at the same time,
we have $\tilde{\mu} \sim \omega_{c}$ so that 
$\tilde{\mu}/\omega_{c}= O(1)$.
Consequently, the direct substitution of the variable $\mu/\omega_{c}$ 
by $\mu_{0}/\omega_{c}$ in the sine arguments of the explicit expression 
for $\tilde{\Omega}$ is no longer valid. Contrary to the 3D metals, 
the exact resolution of the implicit Eq. (2) is needed. 
The consideration of the explicit expression for $\tilde{F}$ as a Fourier 
series, as done in Ref. 5, is now incorrect.
Furthermore, on account of the non negligible oscillating part $\tilde{\mu}$, 
the individual band contributions are also mixed via the nonlinear coupling 
in $\tilde{\Omega}$, so that combination frequencies can be 
produced by both terms in Eq. (1).
Hence, we conclude that the analysis proposed by Alexandrov and Bratkovsky ~\cite{5} for 
explaining the appearance of combination frequencies in 2D multiband metals is mathematically 
incorrect.

Moreover, the mechanism proposed by Alexandrov and Bratkovsky is based on the relation $F= \Omega + \mu N$ which is valid only 
at the thermodynamic limit.
In this limit, the 
thermodynamic quantities evaluated in the canonical ensemble or in the grand 
canonical ensemble are equal for the same given experimental condition. The magnetization oscillations function $\tilde{M}$ 
is then equivalently derived by 
$$\tilde{M}= - \left(\frac{ \partial \tilde{\Omega}}{\partial H} \right)_{\mu}=- \left(\frac{ \partial \tilde{F}}{\partial H} \right)_{N}.$$
 Consequently, the mechanism responsible for the combination frequencies in 
the Fourier spectrum of magnetization oscillations can not a priori depend on 
the way the magnetization is derived, that is to say on the use of a specific 
thermodynamic potential. 

Our following goal is then to point out how the combination frequencies arise by considering directly the expression for the magnetization oscillations in the relevant 
thermodynamic limit.
Exactly like in the 3D Lifshitz-Kosevich derivation~\cite{6}, the 
calculations of magnetization oscillations~\cite{8} are technically easier 
with the grand canonical thermodynamic potential $\Omega$.  Therefore, the 
magnetization oscillations are derived  by way of
\begin{equation}
\tilde{M}=-\left( \frac{\partial \tilde{\Omega}}{\partial H }\right)_{\mu},
\end{equation}
which is a function of the grand canonical variables $\mu$, $V$ and $T$.
For a fixed total number of electrons $N$, the difficulty is still to 
eliminate the variable $\mu$ in the expression for $\tilde{M}$ via the 
condition (3).
As previously stated, the expression (3) for the chemical potential 
oscillations is the key-equation to understand the difference between 2D and 
3D dHvA effects.
Contrary to the 3D case, its exact resolution is needed.
The possible presence of a finite intrinsic reservoir of electrons~\cite{2} 
which reduces the chemical potential oscillations has also to be taken into 
account as an additional parameter.
As a result, in 2D metals, the magnetization oscillations become more or 
less sensitive to the presence of the chemical potential oscillations and 
exhibit significantly different behaviors depending on the presence or not 
of a finite reservoir of electrons~\cite{2,9,10}. 
In 2D multiband metals this higher sensitivity to chemical potential 
oscillations is expressed by the presence of the combination frequencies in 
the Fourier spectrum of magnetization oscillations at very low temperatures, 
as shown numerically by Nakano~\cite{11}. 
Here, our aim is to prove analytically the existence of these combination 
frequencies in the magnetization oscillations.

In 2D one-band metals, the oscillating part $\tilde{\mu}$ and the 
magnetization oscillations $\tilde{M}$ are linked through a simple relation 
of proportionality so that equation (3) and equation (4) can be compiled to give a 
single equation~\cite{9}

\begin{eqnarray}
\frac{\tilde{M}}{M_{0}}=
\frac{2}{\pi}
\sum_{l=1}^{+\infty} 
&&\frac{(-1)^{l+1}}{l}  \sin\left(2 \pi l \frac{\mu_{0}}{\hbar \omega_{c}}+\frac{\pi l}{(1+R)}\frac{\tilde{M}}{M_{0}}\right) \nonumber \\
&&\times \frac{\lambda_{l}}{\sinh 
\lambda_{l}} \cos\left(2 \pi l 
\frac{\mu_{e}H}{\hbar \omega_{c}}\right) \exp\left(-2 \pi l 
\frac{\Gamma}{\hbar \omega_{c}}\right),
\end{eqnarray}
where $\omega_{c}=eH/mc$ is the cyclotron frequency and $m$ the effective mass,
 $M_{0}$ is the magnetization at saturation, 
$\lambda_{l}=2 \pi^{2} l k_{B} T/\hbar \omega_{c}$,  
$\Gamma$ is the width of the Landau levels due to impurity 
scattering (assuming a Lorentzian broadening), and  $\mu_{e}$ is the 
electron's magnetic moment. The dimensionless parameter $R$ measures the
 strengh of the coupling to the background reservoir~\cite{2} (for a more precise definition
 of $R$, see Ref. 9). The chemical potential oscillations are responsible for 
the nonlinearity of Eq. (5). For $R \gg 1$, they are strongly reduced : 
then, formula (5) yields directly the Fourier development of the magnetization  
oscillations in 2D metals for a fixed chemical potential~\cite{1}.

In the multiband case, Eq. (3) and (4) give for $k_{B}T \ll \mu$ and 
$\hbar \omega_{c} \ll \mu$

\begin{eqnarray}
\tilde{M}&=&\sum_{\alpha, l} 
M_{0\alpha} \frac{\mu_{\alpha}}{\mu_{0 \alpha}} A_{\alpha}^{l}
\sin\left(2 \pi l \frac{\mu_{0\alpha}+\tilde{\mu}}{\hbar \omega_{c\alpha}}\right)
,\\
\frac{\tilde{\mu}}{\hbar \omega_{c\alpha}}&=&\frac{1}{2 (1+R)} 
\frac{m_{\alpha}}{m}\sum_{\alpha',l'}
A_{\alpha'}^{l'}\sin\left(2 \pi l' \frac{\mu_{0\alpha'}+
\tilde{\mu}}{\hbar \omega_{c\alpha'}}\right),
\end{eqnarray}
 where $\mu_{\alpha}=\mu_{0\alpha}+\tilde{\mu}$, 
$\mu_{0\alpha}=\mu_{0}-\Delta_{\alpha 0}$ is 
{\em independent of the magnetic field}, 
$M_{0 \alpha}= \rho_{\alpha} \mu_{0\alpha} \hbar \omega_{c\alpha}/H$, 
$\rho_{\alpha}$ is the zero-field density of states per spin projection in 
the band $\alpha$ with the effective mass 
$m_{\alpha}$, $\omega_{c \alpha}=eH/m_{\alpha}c$, 
$m=\sum_{\alpha} m_{\alpha}$, 
$R=\rho_{R}/\sum_{\alpha} \rho_{\alpha}$  (with $\rho_{R}$ the density of
 states of a non-quantized background reservoir),
and
$$A_{\alpha}^{l}=\frac{2}{\pi} \frac{(-1)^{l+1}}{l} 
\frac{\lambda_{l\alpha}}{\sinh 
\lambda_{l\alpha}} \cos\left(2 \pi l 
\frac{\mu_{e}H}{\hbar \omega_{c\alpha}}\right)\exp\left(-2 \pi l 
\frac{\Gamma}{\hbar \omega_{c\alpha}}\right).
$$
We note here that the equations (6) and (7) can not be combined to give a single 
equation as was the case for one band. Nevertheless, the effects of $\tilde{\mu}$ 
on the magnetization oscillations are qualitatively the same.
As for the one-band metal, for $R \gg 1$, the chemical potential oscillations 
are strongly damped and  the Fourier expression with classical frequencies 
$f_{\alpha}=2 \pi c m_{\alpha}\mu_{0\alpha}/e$ is recovered. For a smaller 
value of $R$, the contributions of each band to the magnetization 
oscillations are mixed through the oscillating part of the chemical 
potential $\tilde{\mu}$.  
However, from Eq. (6) and (7), it is not clear whether a Fourier analysis 
for the magnetization oscillations is relevant. 
Let's try to make a Fourier development.
For this purpose, in Eq. (6) and (7) we separate the two different parts 
in the sine arguments : 

\begin{eqnarray}
\tilde{M}=
\sum_{\alpha, l} 
M_{0 \alpha} \frac{\mu_{\alpha}}{\mu_{0 \alpha}} 
A^{l}_{\alpha}&&\left[\sin \left(2 \pi l \frac{\mu_{0\alpha}}{\hbar \omega_{c\alpha}}\right)
\cos \left( 2 \pi l \frac{\tilde{\mu}}{\hbar \omega_{c\alpha}}\right) \right.
 \nonumber \\
&&+ \left. \cos \left(2 \pi l \frac{\mu_{0\alpha}}{\hbar \omega_{c\alpha}}\right) 
\sin \left( 2 \pi l \frac{\tilde{\mu}}{\hbar \omega_{c\alpha}} \right) \right]
,\\
\frac{\tilde{\mu}}{\hbar \omega_{c \alpha}}= \frac{1}{2 (1+R)} \frac{m_{\alpha}}{m}
\sum_{\alpha', l'} 
A^{l'}_{\alpha'}&&\left[\sin \left(2 \pi l' 
\frac{\mu_{0\alpha'}}{\hbar \omega_{c\alpha'}}\right)
\cos \left( 2 \pi l' \frac{\tilde{\mu}}{\hbar \omega_{c\alpha'}}\right) \right.
 \nonumber \\
&&+ \left. \cos \left(2 \pi l' 
\frac{\mu_{0\alpha'}}{\hbar \omega_{c\alpha'}}\right) 
\sin \left( 2 \pi l' \frac{\tilde{\mu}}{\hbar \omega_{c\alpha'}} \right) \right].
\end{eqnarray}
At any finite temperature or impurity scattering, the quantity 
$\tilde{\mu}/\hbar \omega_{c \alpha}$ is strictly less than 
$0.5 (1+R)^{-1} m_{\alpha}/m$ and is reduced further at higher temperature or 
lower magnetic field. Thus, expanding 
$\tilde{\mu}/\hbar \omega_{c \alpha}$ in powers of a temperature or 
impurity reduction factor, Eq. (9) could be solved by iteration.
However, the nonlinearity makes the resolution of this self-consistent 
equation somewhat cumbersome, especially for strong chemical potential 
oscillations $|\tilde{\mu}/\hbar \omega_{c \alpha}| \sim 1/2$.
According to Eq. (8) and (9), it is worth noting that, generally, 
for a fixed number of electrons, a Fourier development of the magnetization 
oscillations may not exist. Nevertheless, in some particular regimes or 
under specific conditions, the description in terms of a Fourier series seems 
possible {\em locally}, that is to say for a finite range of magnetic field 
(which depends on temperature, on the parameter $R$, on the impurity 
broadening $\Gamma$ and also on the electron hopping integral $t$ in quasi-2D 
metals).
Indeed, in the regime of small but non negligible chemical potential 
oscillations $|\tilde{\mu}/\hbar \omega_{c\alpha} |< 1/2\pi $, the 
linearization of Eq. (8) and (9) is possible (for the first 
significant $l$) and yields at first order in chemical potential oscillations

\begin{eqnarray}
\tilde{M}=
\sum_{\alpha, l} M_{0 \alpha} A^{l}_{\alpha}
\left[
\sin\left(2 \pi l \frac{\mu_{0\alpha}}{\hbar \omega_{c\alpha}}\right)
+2 \pi l \frac{\tilde{\mu}}{\hbar \omega_{c \alpha} }
\cos \left(2 \pi l \frac{\mu_{0\alpha}}{\hbar \omega_{c\alpha}}\right) \right],
\end{eqnarray}
where 
\begin{eqnarray}
\frac{\tilde{\mu}}{\hbar \omega_{c \alpha}}=
\frac{1}{2(1+R)} \frac{m_{\alpha}}{m} 
\sum_{\alpha', l'} A^{l'}_{\alpha'}
\sin\left(2 \pi l' \frac{\mu_{0\alpha'}}{\hbar \omega_{c\alpha'}}\right).
\end{eqnarray}
The substitution of (11) into (10) leads to the Fourier series expansion
\begin{eqnarray}
\tilde{M}=&&
\sum_{\alpha,l} M_{0 \alpha}
A_{\alpha}^{l} \sin\left(2 \pi l 
\frac{\mu_{0\alpha}}{\hbar \omega_{c\alpha}}\right) \nonumber \\
&&+
\sum_{\alpha, \alpha',l,l'}\frac{\pi l}{(1+R)} 
\frac{m_{\alpha}}{m} M_{0 \alpha} A^{l}_{\alpha} 
A^{l'}_{\alpha'} \sin\left(2 \pi l' 
\frac{\mu_{0\alpha'}}{\hbar \omega_{c\alpha'}}\right) 
\cos \left(2 \pi l \frac{\mu_{0\alpha}}{\hbar \omega_{c\alpha}}\right) .
\end{eqnarray}
In the presence of several bands, the assumption of small chemical potential 
oscillations is not so restrictive even for $R=0$, since the amplitudes of 
oscillations are reduced by the extra factor $m_{\alpha}/m < 1$. In this 
regime, it is the second term of (12) which is responsible for the presence 
of the combination frequencies $f=l f_{\alpha} \pm l' f_{\alpha'}$. Their 
amplitudes are 
\begin{equation}
\frac{\pi A_{\alpha}^{l} A_{\alpha'}^{l'}}{2(1+R)} 
\left(l \frac{m_{\alpha}}{m} M_{0 \alpha} \pm l' \frac{m_{\alpha'}}{m}
 M_{0 \alpha'} \right).
\end{equation}
At $T=0$, the ratio of the amplitude of the combination frequency 
$f_{\alpha}+f_{\alpha'}$ harmonic to the single band $\alpha$ amplitude 
given by the first term is ($A_{\alpha}^{1}=2/\pi$ without spin-splitting 
and impurity factors)
\begin{equation}
\frac{1}{1+R} \frac{m_{\alpha}}{m} \left(1+\frac{f_{\alpha'}}{f_{\alpha}} \right).
\end{equation}
This value is not incompatible with the condition of linearization. It shows 
that the presence of combination frequencies is significant  if 
 $f_{\alpha}$ and $f_{\alpha'}$ do not have the same order of magnitude. 
In experiments of Shepherd {\em et al.}~\cite{4}, the ratio of band frequencies is of the order of 4, 
which is consistent with this condition for the observation of the combination 
frequencies. In the numerical work of Nakano~\cite{11}, significant 
amplitudes for the combination frequencies are found in the same 
configuration of $f_{\alpha'}/f_{\alpha} \sim 4$.
It means that the 2D multiband metals with individual band frequencies such 
that $f_{\alpha'}/f_{\alpha} \gg 1$ are ideal systems to observe combination 
frequencies in the Fourier spectrum of magnetization oscillations at very 
low temperatures.

Moreover, from Eq. (12) it is seen that the chemical potential oscillations 
modify the amplitudes of the single band frequencies. Indeed, 
the amplitude of the frequency $f=L f_{\alpha}$ consists of the usual 
amplitude (first term in the right-hand side of Eq. (12)) plus an infinity 
of terms produced by the chemical potential oscillations  in the second term 
when $L=l \pm l'$. For example, the ratio of the amplitude of the 
harmonic $L$ (frequency $Lf_{\alpha}$) in the presence of small 
chemical potential oscillations to the usual one is 
\begin{equation}
1+\frac{\pi }{2(1+R)} \frac{m_{\alpha}}{m} \frac{L}{A_{\alpha}^{L}} 
\left( \sum_{l=1}^{L-1} A_{\alpha}^{l} A_{\alpha}^{L-l} - 
\sum_{l=L+1}^{+ \infty}A_{\alpha}^{l}A_{\alpha}^{l-L} \right).
\end{equation}
For the first harmonic $L=1$, this ratio becomes 
\begin{equation}
1-\frac{\pi }{2(1+R)} \frac{m_{\alpha}}{m}\frac{1}{A_{\alpha}^{1}} 
\sum_{l=1}^{+ \infty}A_{\alpha}^{l}A_{\alpha}^{l+1} .
\end{equation}
Using the formula $\sum_{l}1/l(l+1)=3/2$, we find at zero temperature and 
in absence of spin-splitting and impurity scattering : 
\begin{equation}
1+\frac{3}{2} \frac{1}{1+R}\frac{m_{\alpha}}{m}.
\end{equation}
In the regime of small chemical potential oscillations,
 the ratio (17) deviates slightly from unity, so that the ratio of the 
amplitude of the combination frequency $f_{\alpha}+f_{\alpha'}$ to the 
single band $f_{\alpha}$ amplitude [Eq. (14)] is unaffected by this 
correction factor.
However, the expressions (15), (16) and (17) derived with the linearisation 
of the chemical potential oscillations
indicate that in the regime of strong chemical potential oscillations (for 
which numerical calculations of the magnetization oscillations are 
unavoidable) the usual single band amplitudes are also significantly
 affected~\cite{2,10}.

In conclusion, the dHvA effect in 2D multiband metals at a constant number 
of electrons has been investigated analytically, emphasizing the clear difference from 3D metals. In the presence of strong chemical potential 
oscillations, the quantitative description of the magnetization oscillations 
in terms of a Fourier series may break down and has to be done
 numerically. In the regime of small chemical potential oscillations, the 
Fourier analysis is still a good approximation. In 2D multiband metals, these 
oscillations are responsible for the presence of combination frequencies 
which can occur with a significant amplitude if the individual band 
frequencies differ significantly.

I would like to thank J. Villain and S. Roche for stimulating discussions, 
and V.P. Mineev for valuable help.


\begin{thebibliography}{99}
\bibitem{7}
Actually, we should say ``Fourier-like series'' because the magnetization 
oscillations are not strictly speaking periodical with magnetic field : this 
specifies a series in which oscillations are still given by harmonics, but 
with a field dependent amplitude including for example finite temperature or 
impurity scattering effects.
\end{thebibliography}
\end{document}